\documentclass{PoS}

\title{$\pi^+\pi^+$, $K^+ K^+$ and $BB$ Interactions}

\ShortTitle{$\pi^+\pi^+$, $K^+ K^+$ and $BB$}

\author{\speaker{Martin J. Savage}\thanks{I thank the NPLQCD Collaboration for
    producing the results I am presenting. NT@UW-08-19.}\\
        Department of Physics, University of Washington, Seattle, WA98195-1560,
        USA\\
        E-mail: \email{savage@phys.washington.edu}}

\abstract{The most recent calculations of $\pi^+\pi^+$ and  $K^+
  K^+$ scattering by the NPLQCD collaboration using domain-wall valence quarks  on staggered MILC configurations
are presented.  
In addition, a quenched calculation of the potentials between two B-mesons is discussed.}

\FullConference{The XXVI International Symposium on Lattice Field Theory\\
                 July 14-19 2008\\
                 Williamsburg, Virginia, USA}

\begin{document}

\section{Introduction}

Pion-pion scattering at low energies is the simplest and
best-understood hadron-hadron scattering process.  Its simplicity and
tractability follow from the fact that the pions are identified as the
pseudo-Goldstone bosons associated with the spontaneous breaking of
the approximate $SU(2)_L\otimes SU(2)_R$
chiral symmetry of QCD.  For this reason, the
low-momentum interactions of pions are strongly constrained by the
approximate chiral symmetry, more so than other hadrons.  The
scattering lengths for $\pi\pi$ scattering in the s-wave are uniquely
predicted at leading order (LO) in chiral perturbation theory 
($\chi$-PT)~\cite{Weinberg:1966kf}:
\begin{eqnarray}
m_\pi\  a_{\pi\pi}^{I=0} \ = \ 0.1588 \ \ \ \ ; \ \ \ \ m_\pi\  a_{\pi\pi}^{I=2} \ = \
-0.04537 
\ \ \ ,
\label{eq:CA}
\end{eqnarray}
at the physical charged pion mass.
Subleading orders in the chiral expansion of the $\pi\pi$ amplitude
give rise to perturbatively-small deviations from the tree level, and
contain both calculable non-analytic contributions and analytic terms
with new coefficients that are not determined by chiral symmetry
alone~\cite{Gasser:1983yg,Bijnens:1995yn,Bijnens:1997vq}.  In order to
have predictive power at subleading orders, these coefficients must be
obtained from experiment or computed with Lattice QCD.
While the perturbative expansion of the $\pi\pi$
scattering amplitude is expected to converge rapidly, the $KK$ amplitude is
expected to receive sizable contributions from higher orders.
Naive expectations suggest that perturbative corrections to the
KK scattering amplitude are set by $m_K^2/\Lambda_\chi^2$.

Recently, we (NPLQCD) have performed the first $n_f=2+1$ flavor QCD calculation
of the $\pi^+\pi^+$ and $K^+K^+$ scattering 
lengths~\cite{Beane:2005rj,Beane:2007xs,Beane:2007uh}.  The $\pi^+\pi^+$
scattering length has been calculated with percent level precision.
Domain-wall valence quarks were computed on various ensembles of MILC lattices
(staggered sea quarks), and mixed-action chiral perturbation theory~\cite{Chen:2005ab,Chen:2006wf} 
was used to eliminate the leading effects of the finite lattice-spacing.
The results of the lattice calculations are found to be 
consistent with tree-level chiral perturbation theory, even at large pion and
kaon masses, within the uncertainties of the calculations.
We have also performed a quenched calculation of the potentials between two
B-mesons.
As the effective field theory that gives rise to these potentials is the same
as that describing the interactions between nucleons (up to the values of the
counterterms), these potentials provide insight into
the interactions between two or more nucleons.

\section{Hadronic Interactions, the Maiani-Testa Theorem and L\"uscher's
  Method}
\noindent
Extracting hadronic interactions from Lattice QCD calculations is far
more complicated than the determination of the spectrum of stable particles.
This is encapsulated in the
Maiani-Testa theorem~\cite{Maiani:1990ca}, which states that S-matrix elements
cannot be extracted from infinite-volume Euclidean-space Green functions except
at kinematic thresholds~\footnote{An infinite number of infinitely precise
  calculations would allow one to circumvent this theorem.}.
Of course, it is clear from the statement of this theorem how it can be evaded,
one computes Euclidean-space correlation functions at finite volume to extract
S-matrix elements, the formulation of which was known for decades in the
context of non-relativistic quantum mechanics~\cite{Huang:1957im} and extended to quantum field theory
by L\"uscher~\cite{Luscher:1986pf,Luscher:1990ux}.
L\"uscher showed that the energy of two particles in a  finite volume depends
in a calculable way upon their elastic scattering amplitude and their masses
for energies below the inelastic threshold. 
As a concrete example consider $\pi^+\pi^+$ scattering.
A $\pi^+\pi^+$ correlation function 
in the $A_1$ representation of the cubic group~\cite{Mandula:ut} 
(that projects onto the s-wave state in the continuum limit) is
\begin{eqnarray}
C_{\pi^+\pi^+}(p, t) & = & 
\sum_{|{\bf p}|=p}\ 
\sum_{\bf x , y}
e^{i{\bf p}\cdot({\bf x}-{\bf y})} 
\langle \pi^-(t,{\bf x})\ \pi^-(t, {\bf y})\ \pi^+(0, {\bf 0})\ \pi^+(0, {\bf 0})
\rangle
\ \ \ . 
\label{pipi_correlator} 
\end{eqnarray}
In relatively large lattice volumes the energy
difference between the interacting and non-interacting two-meson states
is a small fraction of the total energy, which is dominated by the
masses of the mesons.  In order to extract this energy difference 
the ratio of correlation functions, $G_{\pi^+ \pi^+}(p, t)$, can be formed, 
where
\begin{eqnarray}
G_{\pi^+ \pi^+}(p, t) & \equiv &
\frac{C_{\pi^+\pi^+}(p, t)}{C_{\pi^+}(t) C_{\pi^+}(t)} 
\ \rightarrow \ \sum_{n=0}^\infty\ {\cal A}_n\ e^{-\Delta E_n\ t} 
\  \ ,
\label{ratio_correlator} 
\end{eqnarray}
and the arrow denotes the large-time behavior of $G_{\pi^+ \pi^+}$. 
The single pion correlation function is $ C_{\pi^+}(t)$.
The energy eigenvalue,  $E_n$, and its deviation from the sum of the rest
masses of the particle, $\Delta E_n$, are related to the
center-of-mass momentum $p_n$ by
$
\Delta E_n \  \equiv \ 
E_n\ -\  2m_\pi\ =\ 
\ 2\sqrt{\ p_n^2\ +\ m_\pi^2\ } 
\ -\ 2m_\pi$.
To obtain $p\cot\delta(p)$, where $\delta(p)$ is the phase shift, the
square of the center-of-mass momentum, $p$, is extracted from this
energy shift and inserted
into~\cite{Huang:1957im,Luscher:1986pf,Luscher:1990ux,Hamber:1983vu}
\begin{eqnarray}
p\cot\delta(p) \ =\ {1\over \pi L}\ {\bf
  S}\left(\,\left(\frac{p L}{2\pi}\right)^2\,\right)
\ \ ,
\label{eq:energies}
\end{eqnarray}
which is valid below the inelastic threshold. The regulated three-dimensional sum is~\cite{Beane:2003da}
\begin{eqnarray}
{\bf S}\left(\, x \, \right)\ \equiv \ \sum_{\bf j}^{ |{\bf j}|<\Lambda}
{1\over |{\bf j}|^2-x}\ -\  {4 \pi \Lambda}
\ \ \  ,
\label{eq:Sdefined}
\end{eqnarray}
where the summation is over all triplets of integers ${\bf j}$ such that $|{\bf j}| < \Lambda$ and the
limit $\Lambda\rightarrow\infty$ is implicit.
\begin{figure}[!ht]
\centering                  
\includegraphics*[width=0.5\textwidth]{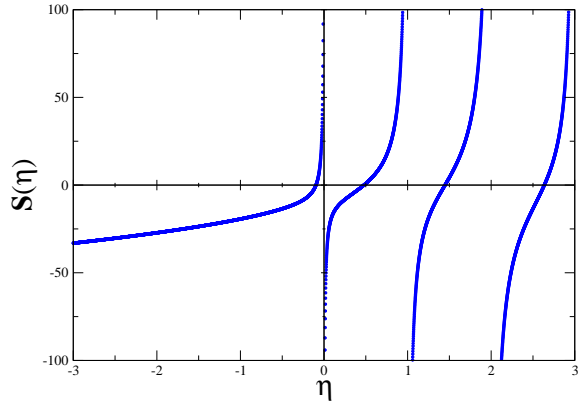}
\caption{The function ${\bf S}({\eta})$ vs.~${\eta}$, defined in
  Eq.~(\protect\ref{eq:Sdefined}), has poles only for ${\eta}\geq 0$.
}
\label{fig:Sfunction}
\end{figure}
Therefore, by measuring the energy-shift, $\Delta E_n$,  of the two particles in the finite lattice
volume, the scattering phase-shift is determined at $\Delta E_n$.

\section{$\pi^+\pi^+$ Scattering}

\begin{figure}[!ht]
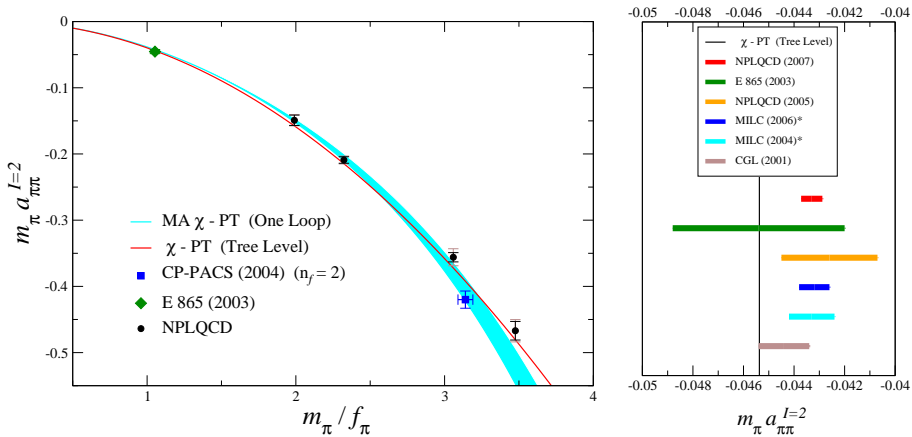

\center
\includegraphics*[width=0.52\textwidth]{figures/fig6.eps}\ \ \ \ 
\includegraphics*[width=0.25\textwidth]{figures/fig7.eps}
\caption{ 
The left panel shows $m_\pi \ a_{\pi\pi}^{I=2}$ vs. $m_\pi/f_\pi$ 
(ovals)~\protect\cite{Beane:2005rj,Beane:2007xs}.
Also shown are the experimental value from
Ref.~\protect\cite{Pislak:2003sv} (diamond) and the lowest quark
mass result of the $n_f=2$ calculation of
CP-PACS~\protect\cite{Yamazaki:2004qb} (square).  The blue band
corresponds to a fit to the lightest three data points 
using the one-loop MA$\chi$-PT formula, and the red line is the tree-level $\chi$-PT result. 
The right panel shows a bar chart of the various determinations of 
the $I=2$ $\pi\pi$ scattering length tabulated in
Table~\protect\ref{tab:vardet}. 
See footnote~\protect\ref{fn:milcstar}.}
\label{fig:barchart}
\end{figure}

\noindent The prediction for the physical value of the $I=2$ $\pi\pi$
scattering length 
from our mixed-action calculation 
is $m_\pi a_{\pi\pi}^{I=2} = -0.04330 \pm 0.00042$~\cite{Beane:2005rj,Beane:2007xs},
which agrees within uncertainties with the (non-lattice) determination
of CGL~\cite{Colangelo:2001df}.
In  Table~\ref{tab:vardet} and Fig.~\ref{fig:barchart} we offer a comparison 
between various determinations~\footnote{The stars on the
MILC results indicate that these are not lattice calculations of the
$I=2$ $\pi\pi$ scattering length but rather a hybrid prediction which
uses MILC's determination of various low-energy constants together
with the Roy equations), and the Roy equation determination of Ref.~\protect\cite{Colangelo:2001df}
(CGL (2001).\label{fn:milcstar}}.
\begin{table}[!ht]
\center
\begin{tabular}{cc}
{} &  $m_\pi\ a_{\pi\pi}^{I=2}$ \\
\hline
$\chi$-PT (Tree Level) & $-0.04438$ \\
NPLQCD (2007) & $-0.04330 \pm 0.00042$ \\
E 865 (2003) & $-0.0454\pm0.0031$ \\
NPLQCD (2005) & $-0.0426\pm 0.0018$ \\
MILC (2006)* & $-0.0432\pm0.0006$ \\
MILC (2004)* & $-0.0433\pm0.0009$ \\
CGL (2001) & $-0.0444 \pm 0.0010$ \\
\end{tabular} 
\caption{A compilation of the various calculations and predictions 
for the $I=2$ $\pi\pi$ scattering length.}
\label{tab:vardet}
\end{table}
%

\section{$K^+ K^+$ Scattering}

\begin{figure}[!t]
\vskip0.5in
\center
\begin{tabular}{c}
\includegraphics[width=0.52\textwidth]{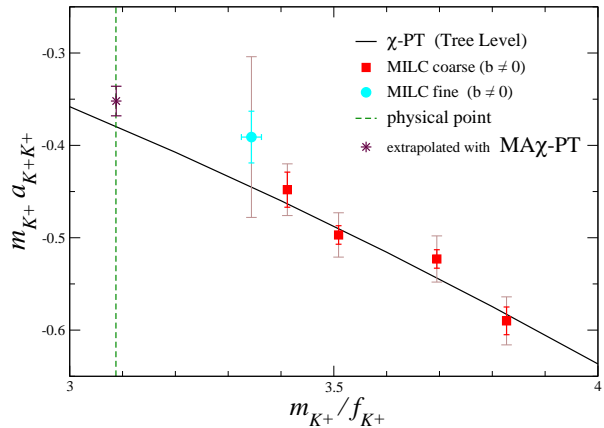} \\
\end{tabular}
\caption{\label{fig:mKaKKdataPhys}
$m_{K^+} a_{K^+ K^+}$ versus $m_{K^+}/f_{K^+}$~\protect\cite{Beane:2007uh}.
The points with error-bars are the results of this lattice calculation (not
extrapolated to the continuum).
The solid curve corresponds to the tree-level prediction of $\chi$-PT. 
The point denoted by a star and its associated uncertainty is the 
chiral extrapolation to the physical meson masses and to the continuum.
}
\end{figure}

\noindent

\noindent 
The $K^+ K^+$ scattering length is calculated in the same way as the 
$\pi^+\pi^+$ scattering length, but requires strange quark
propagators.
The results of the lattice calculation of $K^+ K^+$ scattering 
are extrapolated to the physical values of 
$m_{\pi^+}/f_{K^+} = 0.8731 \pm 0.0096$, 
$m_{K^+}/f_{K^+} = 3.088 \pm 0.018$ and 
$m_\eta / f_{K^+} = 3.425 \pm 0.0019$ assuming isospin
symmetry, and the absence of electromagnetism.
Considering the systematics uncertainties in the chiral extrapolation of the
results shown in Fig.~\ref{fig:mKaKKdataPhys}, along
with the statistical uncertainties, gives
$ m_{K^+} a_{K^+ K^+} \ =\  -0.352 \pm 0.016$~\cite{Beane:2007uh}, 
where the statistical and systematic errors have been combined in quadrature.

\section{BB Potentials}
\noindent
Energy-independent potentials can be rigorously defined and calculated
for systems composed of two (or more) hadrons containing a heavy quark
in the heavy-quark limit, $m_Q\rightarrow\infty$.  This is interesting
for more than academic reasons as the light degrees of freedom
(dof) in the B-meson have the same quantum numbers as the nucleon,
isospin-${1\over 2}$ and spin-${1\over 2}$.  As such, the EFT
describing the interactions between two B-mesons has the same form as
that describing the interactions between two nucleons, but the
counterterms that enter into each EFT are different.  Therefore, a deeper
understanding of the EFT description of nuclear physics can be gained
by Lattice QCD calculations of the potentials between B-mesons.  
We computed the potential between two
B-mesons in the four possible spin-isospin channels (neglecting
$B_d^0-\overline{B}_d^0$ mixing) in relatively small volume DBW2
lattices with $L\sim 1.6~{\rm fm}$, with a pion mass of $m_\pi\sim
403~{\rm MeV}$, and lattice-spacing of $b\sim 0.1~{\rm fm}$~\cite{Detmold:2007wk}.  The
calculation was quenched and the naive Wilson action was used
for the quarks.  At this relatively fine lattice spacing, much finer
than previous calculations, we were able to extract a non-zero potential,
but the small volume meant that the contributions to the potential
from image B-mesons (periodic BC's) were visible.

\begin{figure}[!ht]
\vspace*{4pt}
\center
\includegraphics*[width=0.8\textwidth]{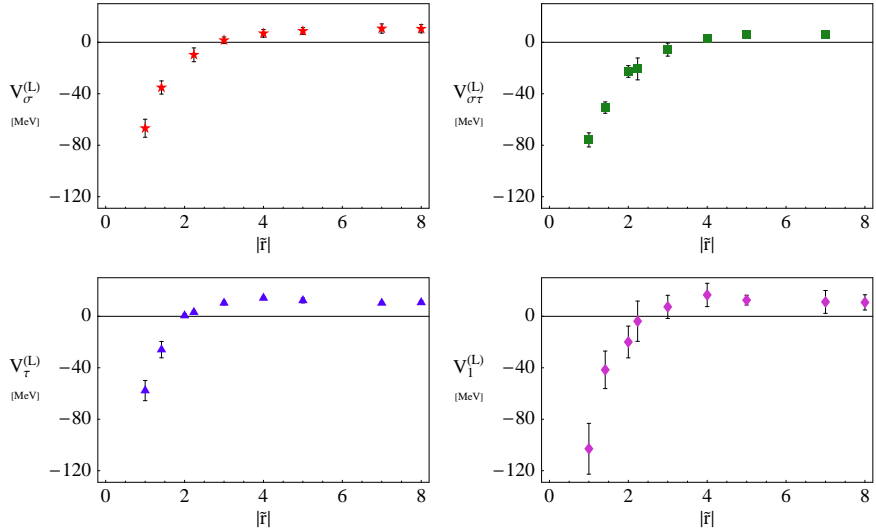}
\caption{The potentials between B-mesons in the finite lattice volume. 
$ V_\sigma^{(L)}, V_\tau^{(L)}, V_{\sigma\tau}^{(L)}$, and $V_1^{(L)} $ 
correspond to the potentials in the exchange-channels with spin-isospin of
$(J,I) = (1,0)$, $(0,1)$, $(1,1)$ and $(0,0)$.
}
\label{fig:BBpots}
\end{figure}

Constructing the t-channel potentials, defined via the quantum
numbers of the exchange particles, in keeping with nuclear physics tradition,
isolated statistical fluctuations into the channel associated with the
``$\sigma$''-meson, leaving the channels with the quantum numbers of the $\pi$,
$\rho$ and $\omega$ with relatively small statistical errors.  The potentials
are shown in Fig.~\ref{fig:BBpots}.

Given the uncertainties in the potentials, and the number of counterterms that
appear in the EFT describing the long- and medium- distance interactions
between the B-mesons, it was possible to make only a parameterization of each
potential beyond the leading light-meson contribution.
Since only the longest range contribution to the potential in each channel
can be identified, we fit our results at large separations, $|{\bf
  r}|>\Lambda_\chi^{-1}$, using the finite-volume versions of the simplified infinite-volume
potentials,
The short distance forms are entirely model
dependent and are the simplest forms that we could find that provide a
reasonable description of the data. Using the measured values and
uncertainties of $m_\pi$ and $m_\rho$ and the physical value of
$f_\pi$ we first determine the light-meson couplings 
by  fitting the finite-volume
potentials at the two largest separations.\footnote{Simple fits using
  the infinite-volume long range behavior were considered in
  Ref.~\cite{Michael:1999nq}.}  The resulting $BB\pi$ coupling is  found
to be $g = 0.57\pm 0.06$.

\section{Conclusions}

I have presented the results of recent calculations by the NPLQCD collaboration
of the $\pi^+\pi^+$ and $K^+ K^+$ scattering lengths, 
and the potentials between two B-mesons.
Percent level precision predictions for the $\pi^+\pi^+$ scattering length were
made possible by
recent theoretical progress in describing mixed-action calculations with chiral
perturbation theory and by the large number of domain-wall propagators ($\sim
2.5\times 10^4$) that were calculated on the coarse MILC lattices ensembles.

The lattice results for meson-meson scattering pose an
interesting puzzle.  
The 
$\pi^+\pi^+$ scattering length tracks the current algebra result up to
pion masses that are expected to be at the edge of the chiral regime
in the two-flavor sector. While in the two flavor theory one expects
fairly good convergence of the chiral expansion and, moreover, one
expects that the effective expansion parameter is small in the channel
with maximal isospin, the lattice calculation clearly imply a cancellation
between chiral logs and counterterms (evaluated at a given scale). 
The same phenomenon occurs in $K^+K^+$ scattering 
(Fig.~\ref{fig:mKaKKdataPhys})
where the chiral expansion is governed by the strange
quark mass and is therefore expected to be more slowly converging. The
$\pi^+K^+$ scattering length exhibits
similar behavior~\cite{Beane:2006gj}. This mysterious
cancellation between chiral logs and counterterms for the meson-meson
scattering lengths begs for an explanation.


\end{document}